\documentclass[12pt]{article}
\usepackage{eurosym}
\usepackage{graphicx}
\usepackage[utf8]{inputenc}
\usepackage[T1]{fontenc}
\usepackage{indentfirst}
\usepackage[margin=0.6in,nomarginpar]{geometry}
\usepackage[final]{hyperref}
\usepackage{amsmath}
\usepackage{hyperref}
\usepackage{cite}
\usepackage{subcaption}
\usepackage{caption}
\usepackage{amssymb}
\usepackage{multirow}

\hypersetup{
colorlinks=true,
linkcolor=blue,
citecolor=blue,
filecolor=magenta,
urlcolor=blue
}

\begin{document}

\title{Black Hole Shadows in Einstein-Bel-Robinson Gravity}
\author{B. Hamil\thanks{%
hamilbilel@gmail.com} \\
Laboratoire de Physique Math\'{e}matique et Subatomique,\\
Facult\'{e} des Sciences Exactes, Universit\'{e} Constantine 1,\\
Constantine, Algeria. \and B. C. L\"{u}tf\"{u}o\u{g}lu\thanks{%
bekir.lutfuoglu@uhk.cz (Corresponding author)} \\
Department of Physics, University of Hradec Kr\'{a}lov\'{e},\\
Rokitansk\'{e}ho 62, 500 03 Hradec Kr\'{a}lov\'{e}, Czechia.}
\date{\today }
\maketitle

\begin{abstract}
Gravity models given by higher-order scalar curvature corrections are believed to bear important consequences. The Einstein-Bel-Robinson gravity, with quartic curvature modification, motivated Sajadi et al to explore static spherically symmetric black hole solutions with perturbative methods. In this paper, inspired by their work, we investigate AdS black hole shadows in EBR gravity.  Moreover, we demonstrate how the gravity parameter alters the energy emission rate. Finally, we handle the same problem in the presence of plasma, since the black holes are thought to be surrounded by a medium that changes the geodesic of photons. 
\end{abstract}


\section{Introduction}
In recent years, research on black holes has been exciting not only for scientists but people everywhere. The existence of black holes, from which even light cannot escape due to their strong gravitational force, is indirectly proven by observing their shadows \cite{M871, SagA1}. In fact, what is observed is the deflection of photons emitted by light sources in the background of the black hole \cite{He}. To be concrete, when the light rays come close to a black hole, photons with low orbital angular momentum are captured by the black hole, while the high orbital angular momentum ones escape. Consequently, an observer at a distance observes a dark zone in the sky, which is known as the black hole's shadow. Naturally, such observations depend on the observer's relative position and the black hole's properties \cite{Pantig2022}. For example, Synge in \cite{Synge66} and later Luminet in \cite{Luminet} showed that the Schwarzschild black hole, a spherically symmetric static black hole, should have a perfectly circular shadow. Not long after the pioneering work of Synge, Bardeen argued that the rotational effects of spinning black holes should distort the perfectly circular shapes of the shadows \cite{Bardeen}. Following the first observation of black hole shadows, in recent years more and more researchers have begun to focus on the theoretical modeling of black hole shadows { \cite{Tsu1, Tsu2, Tsu3, Konoplya, WeiMann, Babar, Kumar, Singh, LiGuo, Zhang21, Thomas1, Das,  GuoWD,  PG1, ZG1, Sunny1, Sunny2, Sunny3, Sunny4, Sunny5, Sunny6, Vir, Vir1,  Biz2023, Biz20231, Du, Akhil, Kara, Ali1, Chowdhuri, Molla, Olmo}}. For a deeper discussion, we refer to two review articles and the references therein \cite{Rev1, Rev2}.


Mathematically, black hole metrics are obtained from solutions of the Einstein equation derived from the Einstein-Hilbert action. In four dimensions the conventional form of the Einstein-Hilbert action is given with the first-order scalar curvature. However, to account for quantum gravity effects the action could be modified. To this end,   many modified gravity theories are given with higher orders of curvature terms in the literature. For example,  Einstein-Gauss-Bonnet gravity is frequently used in the context of black hole physics \cite{MY1, MY2, MY3, MY10, MY11}, the first time by Boulware and Deser  \cite{Boulware},  to discuss the thermal and optical properties \cite{Kumar1, Ul, Heydar1, SS1, MY4, MY5, MY6, MY7, MY8, MY9}. 

The Starobinsky gravity model \cite{Starobinsky}, one of the first f(R) models, is perfectly compatible with cosmological inflation \cite{Pozdeeva}. In 2022, Ketov extended this model perturbatively with quartic curvature terms that are structurally connected to the Bel-Robinson tensor \cite{Bel, Robinson} in four dimensions \cite{Ketov}. Then, Delgado et al derived the Schwarzschild-type black hole metric \cite{Delgado} and examined its thermal properties. Shortly after, Belhaj et al presented its optical properties \cite{Belhaj}. Meanwhile, Sajadi et al explored the static black hole solution in anti-de-Sitter (AdS) spacetime in the Einstein-Bel-Robinson (EBR) gravity model and discussed the black hole's thermodynamics in the presence of first-order small perturbative parameter \cite{Sajadi}. 

{ AdS black holes are of interest because of their thermodynamic stability properties. Moreover, AdS black hole thermodynamics exhibits various phase transitions. A primary interesting example is the thermal radiation/black hole first-order phase transition observed for Schwarzschild-AdS black hole spacetimes. The asymptotic behavior of AdS spacetime allows for well-defined boundary conditions, making the thermodynamics of AdS black holes more tractable. We also observe that due to the AdS/CFT correspondence and certain conformal field theories, there has been a revival of interest in the physics of asymptotically AdS black holes in recent years. On the other hand, the} EBR model differs from others in that it is a ghost-free model and deserves a more in-depth examination. By considering the spherical static solution of Sajadi et al, we intend to reveal the optical properties of the AdS black hole. To this end, we prepare the manuscript in the following form: in Sec. \ref{sec2}, at first we review the EBR gravity briefly and give the mathematical base of the AdS black hole. Then, we study their shadows. In Sec. \ref{sec3}, we investigate the energy emission rate. Then, in Sec. \ref{sec4}, we revisit the optical properties of the AdS black hole physics in EBR gravity in the presence of a plasma background. Finally, we conclude the manuscript in the last section.

\section{AdS black hole in EBR gravity and its shadow} \label{sec2}

In \cite{Sajadi}, Sajadi et al introduced the action of the EBR gravity 
in four dimensions as 
\begin{eqnarray}
    \mathcal{S}&=&\frac{1}{16\pi G}\int d^4x \sqrt{-g} \left(R -2 \Lambda-\beta\left(\mathcal{P}^2-\mathcal{G}^2\right) \right),
\end{eqnarray}
where $g$, $R$, and $\beta$  denote the metric, the scalar curvature, and the stringy coupling constant, respectively. Here, the cosmological constant relates to the curvature radius of the AdS background via $\Lambda=-3 / \ell^2$, however, we prefer to use its given definition with pressure, $p$, in the context of the black hole chemistry, $\Lambda=-8\pi p$,  \cite{Mannchem}. The other quantities are the Euler and Pontryagin topological densities of the form of 
\begin{eqnarray}
    \mathcal{P}&=& \frac{1}{2}\sqrt{-g}\,\epsilon_{\mu\nu\rho\sigma}\,R^{\rho\sigma}_{\alpha\beta}\,R^{\mu\nu\alpha\beta}, \\
    \mathcal{G}&=& R^2-4 R_{\mu\nu} R^{\mu\nu}+ R_{\mu\nu\rho\sigma} R^{\mu\nu\rho\sigma},
\end{eqnarray}
which arise from the Bel-Robinson tensor square. After taking the variation of the action with respect to the metric, Sadaji et al handled the following line element 
\begin{eqnarray}
ds^2=-f\left( r\right) dt^{2}+\frac{1}{f\left( r\right) }dr^{2}+r^{2}d\Omega
\end{eqnarray}
in the equation of motion. After straightforward calculation of the field equations, they obtained the lapse function as 
\begin{eqnarray}
f\left( r\right) &=& 1-\frac{2m}{r}+\frac{1664\beta m^{4}}{r^{10}}+\frac{3072\pi \beta m^{3}p}{r^{7}}-\frac{576\beta m^{3}}{r^{9}}-\frac{40960\pi^{2}\beta m^{2}p^{2}}{3r^{4}}-\frac{12288\pi \beta m^{2}p}{5r^{6}} \nonumber\\&&+\frac{8\pi pr^{2}\left( 9-16384\pi ^{3}\beta p^{3}\right) }{27}+\mathcal{O}(\beta^2).
\end{eqnarray}
We note that in the absence of the coupling constant, the lapse function takes its conventional form.

{ In black hole physics, the roots of the lapse function are used to determine the horizon of the black hole. In the problem we are considering, we see that the coupling constant plays an important role in the values of the horizon and it can be used to determine the critical value of the $\beta $ parameter. To illustrate this, in Fig. \ref{lapsefunc} we plot the lapse function versus the radial coordinate for various $\beta$ parameter values.
\begin{figure}[tbh!]
\centering
\includegraphics[scale=1]{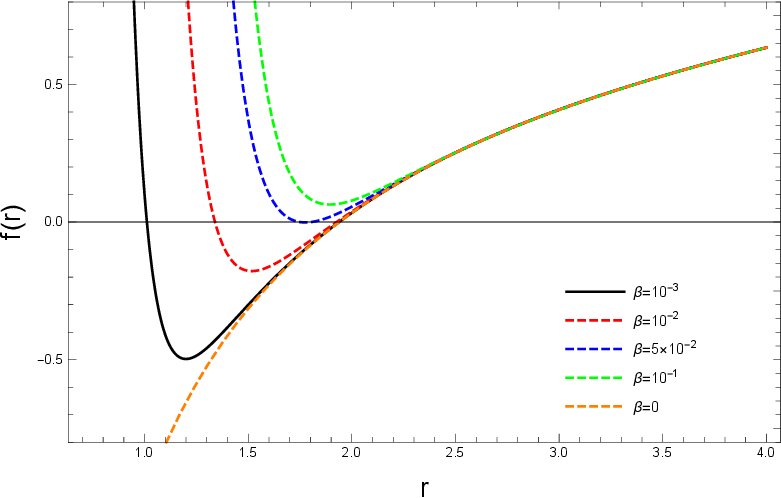}
\caption{The lapse function versus the radial coordinate for $m=1$ and $p=10^{-3}$.}
\label{lapsefunc}
\end{figure}

We observe that depending on the strength of the coupling constant, either two or no horizon can occur. We see that for a critical beta parameter value only one horizon forms. 
}

Now, let us draw your attention to the derivation of the AdS black hole shadow in the context of the EBR gravity.  To this end, we consider the Lagrangian 
\begin{eqnarray}
  \mathcal{L}=\frac{1}{2}g_{\mu \nu }\dot{x}^{\mu }\dot{x}^{\nu }=\frac{1}{2}\left[ -f\left( r\right) \dot{t}^{2}+\frac{1}{f\left( r\right) }\dot{r}^{2}+r^{2}\dot{\theta}^{2}+r^{2}\sin ^{2}\theta \dot{\phi}^{2}\right] ,  
\end{eqnarray}
where the dot represents the derivative with respect to affine parameter $\tau $ and $g_{\mu \nu }$ denotes the metric tensor. By implementing the Euler-Lagrange equation, we easily deduce the following constants of motion 
\begin{eqnarray}
E&=& {\bigg(}1-\frac{2m}{r}+\frac{8\pi pr^{2}\left( 9-16384\pi ^{3}\beta
p^{3}\right) }{27}+\frac{1664\beta m^{4}}{r^{10}}+\frac{3072\pi \beta m^{3}p%
}{r^{7}}-\frac{576\beta m^{3}}{r^{9}} \nonumber \\
&&-\frac{40960\pi ^{2}\beta m^{2}p^{2}}{%
3r^{4}}-\frac{12288\pi \beta m^{2}p}{5r^{6}}{\bigg)}\dot{t},\\
P_{\phi }&=&L=r^{2}\sin ^{2}\theta \dot{\phi}.
\end{eqnarray}
%
Here, $E$ and $L$ illustrate the test particle's energy and angular momentum,  respectively. The next step involves acquiring the geodesic form of the particle. This can be achieved by using the Hamilton-Jacobi equation
\begin{equation}
\frac{\partial S}{\partial \tau }+H=0,  \label{hj}
\end{equation}%
where $S$ and $H$ denote the Jacobi action and the Hamiltonian, respectively.
\begin{equation}
H=\frac{1}{2}g^{\mu \nu }p_{\mu }p_{\nu }.
\end{equation}%
By following the relation between the Jacobi action and the momentum, $p_{\mu }$, 
\begin{equation}
\frac{\partial S}{\partial x^{\mu }}=p_{\mu }, 
\end{equation}%
and by employing the separation method based on Carter constants \cite{Carter},  
\begin{equation}
S=\frac{\Tilde{m}^{2}}{2}\tau -Et+L\phi +S_{\theta }\left( \theta \right)
+S_{r}\left( r\right) ,  \label{ss}
\end{equation}
we get
\begin{equation}
\frac{E^{2}}{f\left( r\right) }-f\left( r\right) \left( \frac{\partial S_{r} }{\partial r}\right) ^{2}-\frac{1}{r^{2}}\left( \frac{L^{2}}{\sin ^{2}\theta}+\mathcal{K-}L^{2}\cot ^{2}\theta \right) -\frac{1}{r^{2}}\left( \left(\frac{\partial S_{\theta }}{\partial \theta }\right) ^{2}-\mathcal{K}+L^{2}\cot ^{2}\theta \right) =0. \label{eq}
\end{equation}%
Here, $\mathcal{K}$ stands for the Carter's constant. It is worth emphasizing that we have to equalize the mass, $\Tilde{m}$, to zero, since the considered particle is a photon.  
%
%
Next, we utilize the relations, $\frac{\partial S_{\theta }}{\partial \theta }=p_{\theta }$, and,   $\frac{\partial S_{r}}{\partial r}=p_{r}$. We get
\begin{eqnarray}
\frac{\partial S_{\theta }}{\partial \theta }&=&r^{2}\frac{\partial \theta }{%
\partial \tau },  \label{teta} \\
\frac{\partial S_{r}}{\partial r}&=&\frac{1}{f\left( r\right) }\frac{\partial r%
}{\partial \tau }.  \label{eqr}
\end{eqnarray}
By embedding Eqs. (\ref{teta}) and (\ref{eqr}) into Eq. (\ref{eq}), we recast it as two separate equations:
\begin{eqnarray}
r^{2}\left( \frac{\partial S_{r}}{\partial r}\right) ^{2}&=&r^{2}\frac{E^{2}}{%
f^{2}\left( r\right) }-\frac{\left( L^{2}+\mathcal{K}\right) }{f\left(
r\right) }, \\
\left( \frac{\partial S_{\theta }}{\partial \theta }\right) ^{2}&=&\mathcal{K}%
-L^{2}\cot ^{2}\theta .
\end{eqnarray}%
Then, we use the canonically conjugate momentum definition to express the complete set of  equations of motion
\begin{eqnarray}
r^{2}\frac{\partial \theta }{\partial \tau }&=&\pm \sqrt{\Theta }\sqrt{%
\mathcal{K}-L^{2}\cot ^{2}\theta }, \\
r^{2}\dot{r}&=&\pm \sqrt{\mathcal{R}}\sqrt{r^{4}E^{2}-r^{2}f\left( r\right)
\left( L^{2}+\mathcal{K}\right) }, \label{rad}
\end{eqnarray}
where
\begin{eqnarray}
\Theta &=& \mathcal{K}-L^{2}\cot ^{2}\theta ,
\end{eqnarray}%
and
\begin{eqnarray}
\mathcal{R}\left( r\right) &=& r^{4}E^{2}-r^{2}f\left( r\right)
\left( L^{2}+\mathcal{K}\right) .
\end{eqnarray}%
Here, the symbols $"+"$ and $"-"$ correspond to the motion of photons in the outgoing and ingoing radial directions, respectively. 

\begin{figure}[htbh!]
\begin{minipage}[t]{0.5\textwidth}
        \centering
        \includegraphics[width=\textwidth]{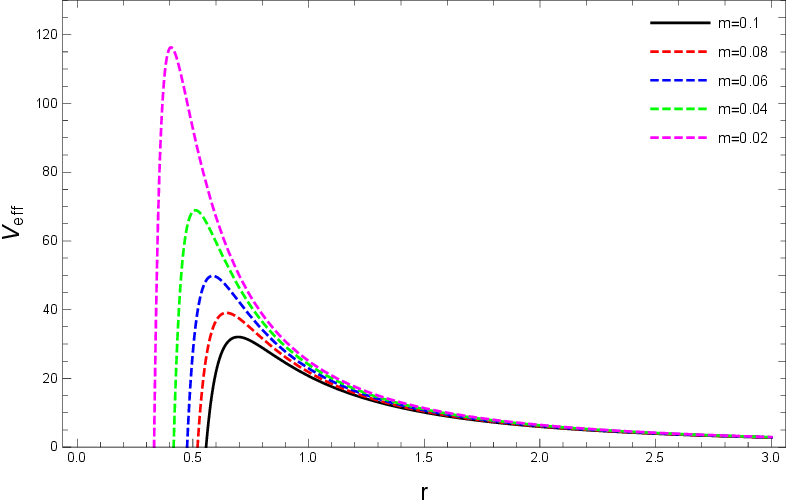}
       \subcaption{ $ \beta=0.01$, and $p=0.005$.}\label{fig:pc}
   \end{minipage}%
\begin{minipage}[t]{0.5\textwidth}
        \centering
        \includegraphics[width=\textwidth]{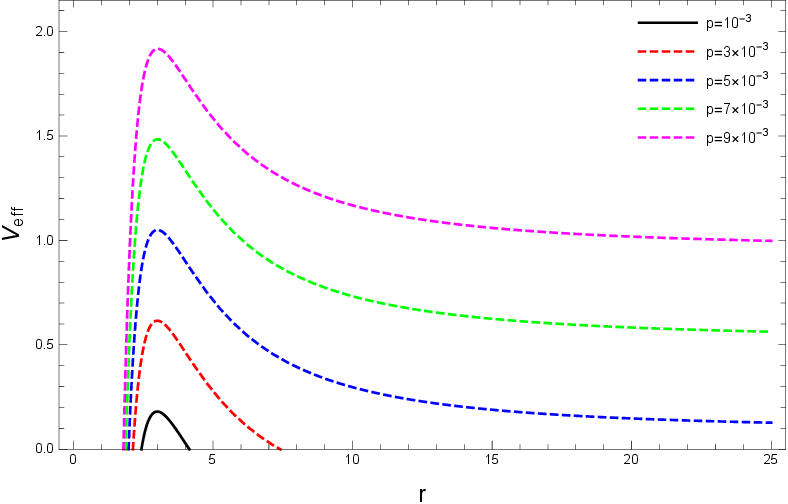}
         \subcaption{ $ m=1$, and $\beta=0.01$.}\label{fig:pd}
   \end{minipage}
\caption{The variation of the effective potential as a function of the radial coordinate for $\protect\beta =0.01$, $L=5$, $\mathcal{K=}1$, and $E=1.$}
\label{vef}
\end{figure}

\begin{figure}[htb!]
\begin{minipage}[t]{0.5\textwidth}
        \centering
        \includegraphics[width=\textwidth]{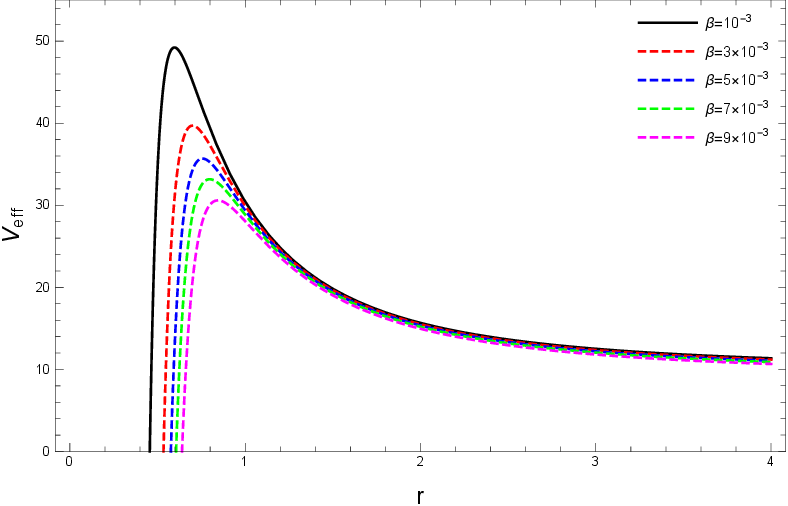}

       \subcaption{ $ m=1$, and $p=0.05$.}\label{fig:va}
   \end{minipage}%
\begin{minipage}[t]{0.5\textwidth}
        \centering
        \includegraphics[width=\textwidth]{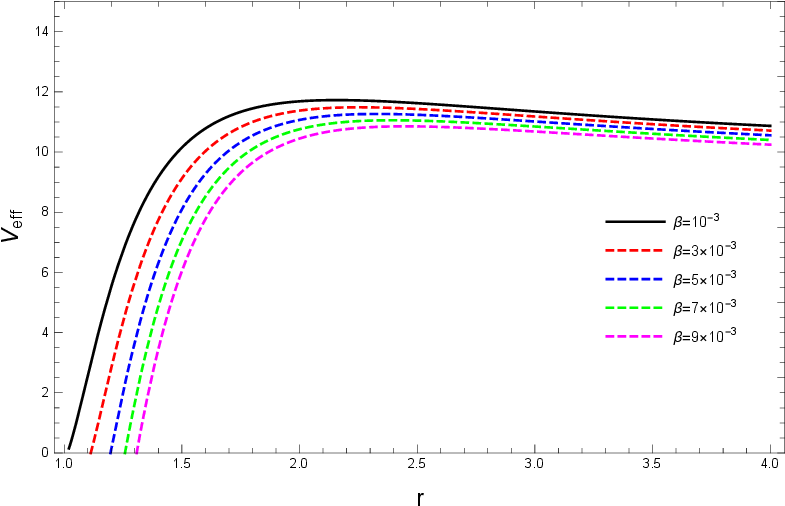}
         \subcaption{ $ m=0.5$, and $p=0.05$.}\label{fig:vb}
   \end{minipage}
\caption{Variation of the effective potential with respect to distance $r$ by
considering different values of $\protect\beta $, we have used $p=0.05$, $L=5$, $\mathcal{K=}1$, and $E=1.$}
\label{figvef}
\end{figure}

It is known that the presence of unstable null circular orbits could indicate the shadow boundary of black holes. To determine those orbits, we have to re-express the radial null geodesic equation in the following manner:
\begin{eqnarray}
\left( \frac{\partial r}{\partial \tau }\right) ^{2}+V_{eff}\left( r\right)
=0.
\end{eqnarray}
Here, the effective potential of the radial motion, $V_{eff}\left( r\right) $,  reads:
\begin{eqnarray}
V_{eff}\left( r\right) =\frac{f\left( r\right) }{r^{2}}\left( L^{2}+\mathcal{%
K}\right) -E^{2}. \label{Vef}
\end{eqnarray}
It is important to mention that the resulting effective potential matches perfectly with the Schwarzschild scenario in the zero limit of $\beta$ and $p$. We think it necessary to briefly investigate the behavior of the effective potential just before obtaining the optical properties. To this end, in Fig. \ref{vef} and Fig. \ref{figvef} we present the plots of the effective potential function versus radial distance for various values of the parameters $m$, $p$, and $\beta$.

In particular, Fig. \ref{vef} demonstrates how the mass and the pressure affect the effective potential. The peaks in these graphs reveal the presence of unstable circular photon orbits. Here, we observe that the peaks are shifting to the left with decreasing values as the BH mass increases for fixed values of $\beta $ and $p$. In addition, when $m$ remains fixed, the maxima are also increasing and shifting to the left as the black hole pressure increases. { Another important remark is that for some pressure values i.e. $p=10^{-3}$, and $p=3\times 10^{-3}$, the effective potential becomes negative for larger $r$ values. These cases show that the test particle can have bound orbits, meaning it cannot reach infinity.}

Fig. \ref{figvef} presents the influence of the EBR parameter on the effective potential. We observe that with an increase in the EBR parameter decreases the peak value of the potential decreases significantly. However, we do not see any impact of this parameter on the photon sphere size of the black hole since the peak position of the effective potential changes its position with respect to $r$.

Next, we impose the conditions \cite{Anish}
\begin{equation}
\left. V_{eff}\left( r\right) \right\vert _{r=r_{p}}=\left. \frac{{d}
V_{eff}\left( r\right) }{{d} r}\right\vert _{r=r_{p}}=0,\text{ \ }
\label{ve}
\end{equation}%
or%
\begin{equation}
\left. \mathcal{R}\left( r\right) \right\vert _{r=r_{p}}=\left. \frac{%
{d}\mathcal{R}\left( r\right) }{{d}r}\right\vert _{r=r_{p}}=0,
\label{ver}
\end{equation}%
to identify the unstable circular orbits, where $r_{p}$ signifies the radius of the photon sphere.
Using { the set of equations given in} Eqs. (\ref{ve}) and  (\ref{ver}), we get
\begin{equation}
2-\frac{6m}{{ r_p}}+\frac{19968\beta m^{4}}{{ r_p}^{10}}+\frac{576\beta m^{3}\left(
48\pi p{ r_p}^{2}-11\right) }{{ r_p}^{9}}-\frac{16384\pi \beta m^{2}p\left( 25\pi
p{ r_p}^{2}+6\right) }{5{ r_p}^{6}}=0,  \label{psp}
\end{equation}
Alternatively don't forget the label
\begin{equation}
2-\frac{6m}{{ r_p}}+ \beta \frac{64 m^2}{r_p^6}\bigg[\frac{312m^2}{r_p^4}+\frac{9m}{r_p^3}\left(
48\pi p{ r_p}^{2}-11\right)-\frac{256\pi p }{5} \left( 25\pi
p{ r_p}^{2}+6\right)\bigg]=0, 
\end{equation}
and then, with the help of two impact parameter definitions
\begin{eqnarray}
  \zeta =\frac{L}{E}, \quad \quad
  \eta =\frac{\mathcal{K}}{E^{2}},
\end{eqnarray}
we find
\begin{equation}
\eta +\zeta ^{2}=\frac{4r_{p}^{2}}{2f\left( r_{p}\right) + { r_p} f^{\prime }\left(
r_{p}\right) }.  \label{con}
\end{equation}
We notice that Eq. (\ref{psp}) can not be solved analytically, therefore we use numerical methods. Since the EBR gravity bears an extra parameter on the solutions, we have to assign different values to $\beta$. Keeping this fact in mind, we obtain numerical solutions for the photon sphere radius and present the computed values of $\eta+\zeta ^{2}$ in Tables \ref{tab1}, \ref{tab2} and \ref{tab3}, respectively. 

\begin{table}[htb!]
\center%
\begin{tabular}{l|ll|ll}
\hline
\multirow{2}{*}{} & \multicolumn{2}{|c|}{$m=1,\beta =0.01$} &
\multicolumn{2}{|c}{$m=0.5,\beta =0.01$} \\ \hline
$p$ & $r_{p}$ & $\eta +\zeta ^{2}$ & $r_{p}$ & $\eta +\zeta ^{2}$ \\
\hline\hline
$10^{-3}$ & 3.00058 & 22.0241 & 1.49757 & 6.42059 \\
$3\times 10^{-3}$ & 3.00312 & 16.095 & 1.51222 & 5.83382 \\
$5\times 10^{-3}$ & 3.00683 & 12.6838 & 1.52793 & 5.34949 \\
$7\times 10^{-3}$ & 3.01169 & 10.4679 & 1.54447 & 4.94251 \\
$9\times 10^{-3}$ & 3.01764 & 8.91302 & 1.56162 & 4.59533 \\ \hline\hline
\end{tabular}%
\caption{The values of photon radius, $r_{p}$, and impact parameters $%
\protect\eta +\protect\xi ^{2}$, for different values of pressure.}
\label{tab1}
\end{table}

\begin{figure}[htb!]
\begin{minipage}[t]{0.5\textwidth}
        \centering
        \includegraphics[width=\textwidth]{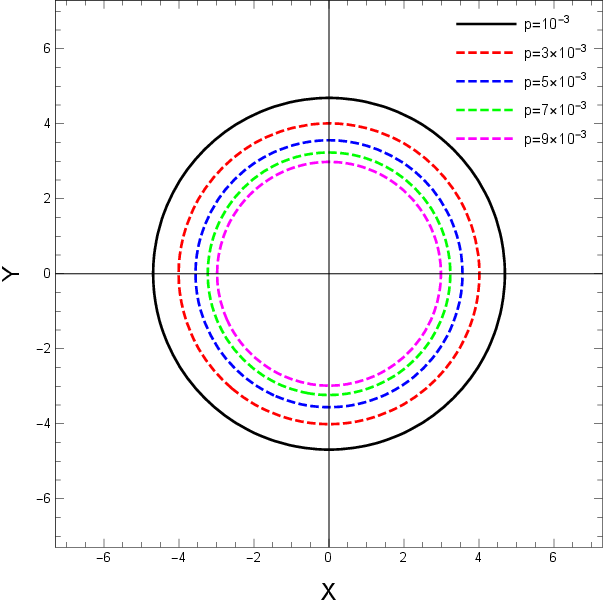}

       \subcaption{ $m=1$, and $\beta=0.01$.}\label{fig:pf}
   \end{minipage}%
\begin{minipage}[t]{0.5\textwidth}
        \centering
        \includegraphics[width=\textwidth]{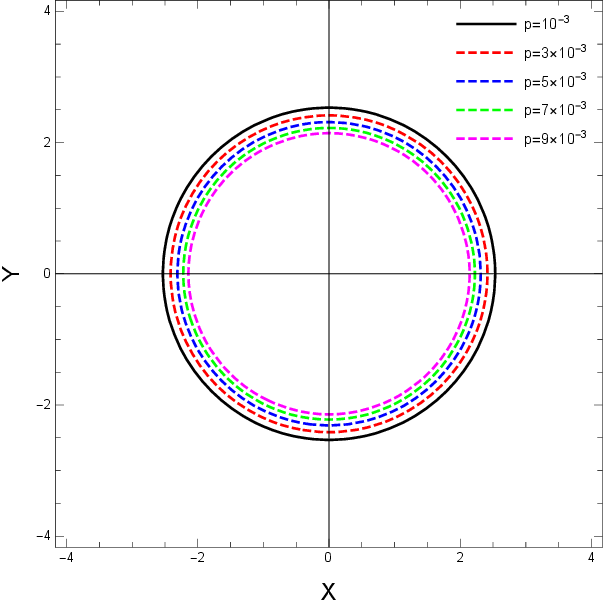}
         \subcaption{ $ m=0.5$, and $\beta=0.01$.}\label{fig:pg}
   \end{minipage}
\caption{Impact of $p$ on the shadow shape within the equatorial plane.}
\label{figsh}
\end{figure}

\begin{figure}[htb!]
\begin{minipage}[t]{0.5\textwidth}
        \centering
        \includegraphics[width=\textwidth]{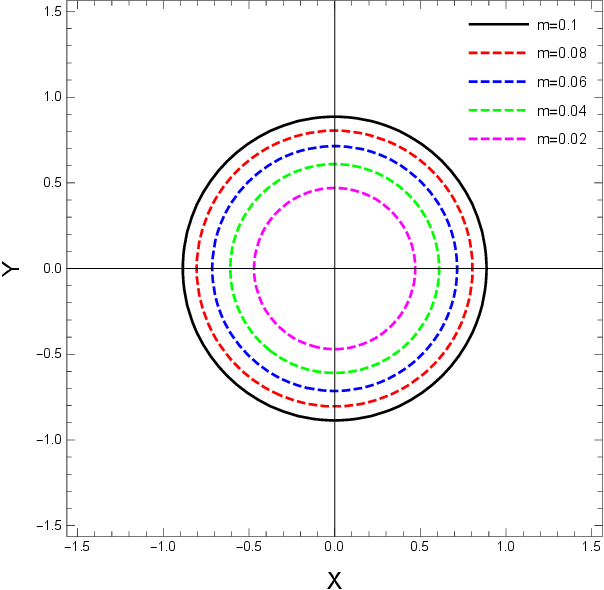}

       \subcaption{ $p=0.005$, and $\beta=0.01$.}\label{fig:ph}
   \end{minipage}%
\begin{minipage}[t]{0.5\textwidth}
        \centering
        \includegraphics[width=\textwidth]{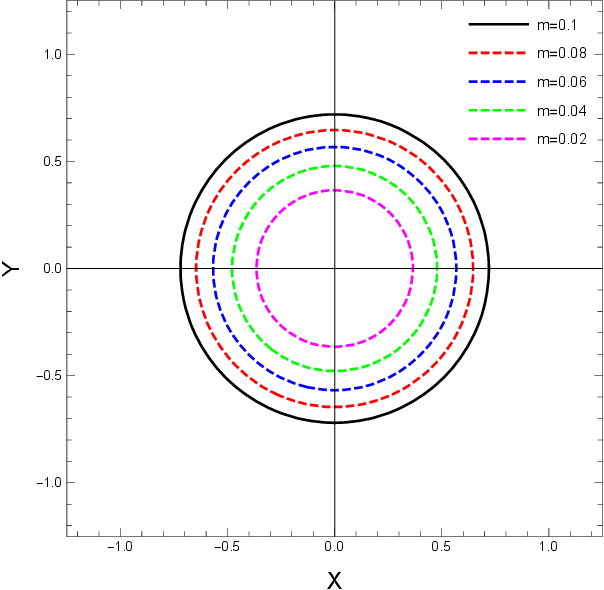}
         \subcaption{ $ p=0.005$, and $\beta=0.001$.}\label{fig:pk}
   \end{minipage}
\caption{Impact of $m$ on the shadow shape within the equatorial plane.}
\label{figsha}
\end{figure}

\begin{figure}[htb!]
\begin{minipage}[t]{0.5\textwidth}
        \centering
        \includegraphics[width=\textwidth]{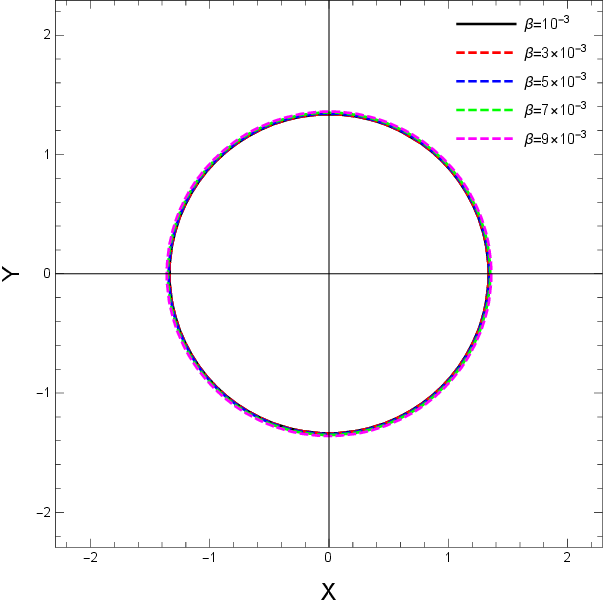}

       \subcaption{ $m=0.4$, and $p=0.05$.}\label{fig:plf}
   \end{minipage}%
\begin{minipage}[t]{0.5\textwidth}
        \centering
        \includegraphics[width=\textwidth]{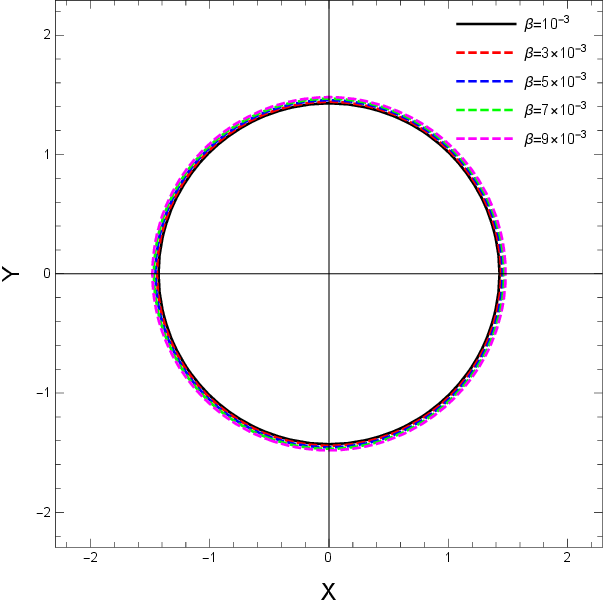}
         \subcaption{ $m=0.7$, and $p=0.05$.}\label{fig:plg}
   \end{minipage}
\caption{Impact of $\beta$ on the shadow shape within the equatorial plane.}
\label{figshadw}
\end{figure}

\begin{table}[htb!]
\center%
\begin{tabular}{l|ll|ll}
\hline
\multirow{2}{*}{} & \multicolumn{2}{|c|}{$p=0.005,\beta =0.01$} &
\multicolumn{2}{|c}{$p=0.005,\beta =0.001$} \\ \hline
$m$ & $r_{p}$ & $\eta +\zeta ^{2}$ & $r_{p}$ & $\eta +\zeta ^{2}$ \\
\hline\hline
$0.1$ & 0.695426 & 0.786735 & 0.531083 & 0.51896 \\
$0.08$ & 0.645059 & 0.648966 & 0.493188 & 0.418694 \\
$0.06$ & 0.585656 & 0.511668 & 0.448215 & 0.32319 \\
$0.04$ & 0.511282 & 0.371854 & 0.391641 & 0.229921 \\
$0.02$ & 0.405547 & 0.221649 & 0.310908 & 0.134011 \\ \hline\hline
\end{tabular}%
\caption{The values of photon radius, $r_{p}$, and impact parameters $%
\protect\eta +\protect\xi ^{2}$, for different values of the black hole
mass. }
\label{tab2}
\end{table}
\begin{table}[htb!]
\center%
\begin{tabular}{l|ll|ll}
\hline
\multirow{2}{*}{} & \multicolumn{2}{|c|}{$p=0.05,m=0.4$} &
\multicolumn{2}{|c}{$p=0.05,m=0.7$} \\ \hline
$\beta $ & $r_{p}$ & $\eta +\zeta ^{2}$ & $r_{p}$ & $\eta +\zeta ^{2}$ \\
\hline\hline
$10^{-3}$ & 1.30949 & 1.79214 & 2.15802 & 2.04302 \\
$3\times 10^{-3}$ & 1.43968 & 1.79292 & 2.2522 & 2.08211 \\
$5\times 10^{-3}$ & 1.52702 & 1.80692 & 2.32839 & 2.11966 \\
$7\times 10^{-3}$ & 1.59474 & 1.82763 & 2.39318 & 2.15654 \\
$9\times 10^{-3}$ & 1.65085 & 1.85168 & 2.44998 & 2.19322 \\ \hline\hline
\end{tabular}%
\caption{The values of photon radius, $r_{p}$, and impact parameters $%
\protect\eta +\protect\xi ^{2}$ for different values of $\protect%
\beta $.}
\label{tab3}
\end{table}
In \cite{Papnoi}, authors proposed to use the celestial coordinates, $X$ and $Y$, to visualize the shadow on the observer's frame. Following that suggestion, we employ the celestial coordinates defined in the following manner,
\begin{eqnarray}
X&=&\lim_{r_{o}\rightarrow \infty }\bigg(-r_{o}^{2}\sin \theta _{o}\frac{d\phi }{dr}\bigg), \\
Y&=&\lim_{r_{o}\rightarrow \infty }\bigg(r_{o}^{2}\frac{d\theta }{dr}\bigg).
\end{eqnarray}
Here, $r_{0}$ and $\theta _{o}$ represent the distance and the angle between the shadow and the observer, respectively. In the case of null geodesic, we get
\begin{eqnarray}
X&=&-\frac{\zeta }{\sin \theta _{o}},  \label{x} \\
Y&=&\pm \sqrt{\eta -\zeta ^{2}\cot ^{2}\theta _{o}}.  \label{y}
\end{eqnarray}
Then, we consider the observer on the equatorial hyperplane and we take $\theta _{o}=\pi /2$. So that Eqs.(\ref{x}) and (\ref{y}) reduce to the following form
\begin{eqnarray}
X&=&-\zeta , \\
Y&=&\pm \sqrt{\eta }.
\end{eqnarray}%
Consequently, Eq.(\ref{con}) becomes 
\begin{equation}
X^{2}+Y^{2}=\eta +\zeta ^{2}=\frac{4r_{p}^{2}}{2f\left( r_{p}\right)
+r_{p} f^{\prime }\left( r_{p}\right) }=R_{s}^{2}.
\end{equation}
{ With the help of a serial expansion up to the first order of $\beta$, the shadow radius, $R_{s}$, reads:
\small
\begin{eqnarray}
R_{s}^{2} &=&\frac{6r_{p}^{2}}{3r_{p}-3m+16\pi pr_{p}^{3}} \nonumber \\
&+&\frac{64\beta \left( 28080m^{4}-8505m^{3}r_{p}+32400\pi
m^{3}pr_{p}^{3}-57600\pi ^{2}m^{2}p^{2}r_{p}^{6}-20736\pi
m^{2}pr_{p}^{4}+40960\pi ^{4}p^{4}r_{p}^{12}\right) }{15r^{7}\left(
3r_{p}-3m+16\pi pr_{p}^{3}\right) ^{2}}. \,\,\,\,\,
\end{eqnarray}
\normalsize
In Figs. \ref{figsh}-\ref{figshadw}, we illustrate the dependence of the shadow radius on the parameters $p$, $m$, and $\beta$. }

{In particular, in Fig. \ref{figsh}, we show} the stereographic projection of the shadow for different values of the black hole pressure. We see that with an increase in the pressure, the black hole's shadow radius decreases. 

\newpage
{ We then demonstrate in Fig. \ref{figsha} that the increase in mass of the black hole causes an increase in its shadow radius. We note that this characteristic change} remains the same for different values of the $\beta$ parameter.

\begin{figure}[htb!]
\begin{minipage}[t]{0.35\textwidth}
        \centering
        \includegraphics[width=\textwidth]{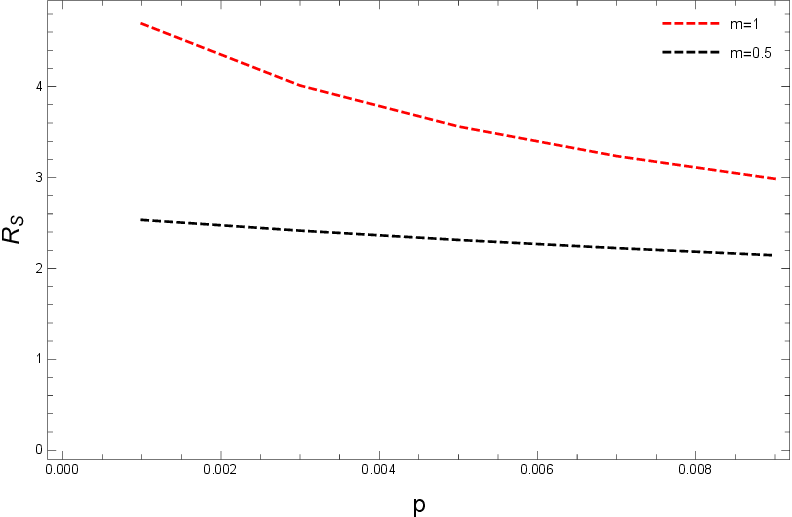}

       \subcaption{ $\beta=0.01$.}\label{fig:Ra}
   \end{minipage}%
\begin{minipage}[t]{0.35\textwidth}
        \centering
        \includegraphics[width=\textwidth]{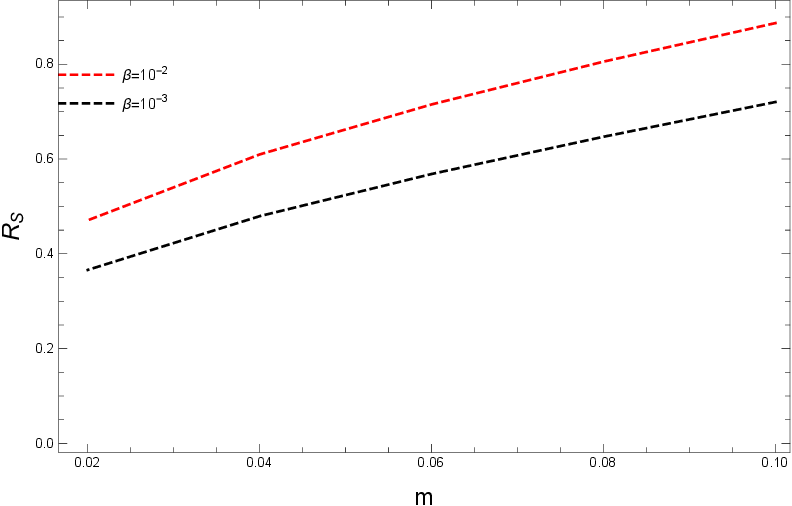}
         \subcaption{ $ p=0.005$.}\label{fig:Rb}
   \end{minipage}%
\begin{minipage}[t]{0.35\textwidth}
        \centering
        \includegraphics[width=\textwidth]{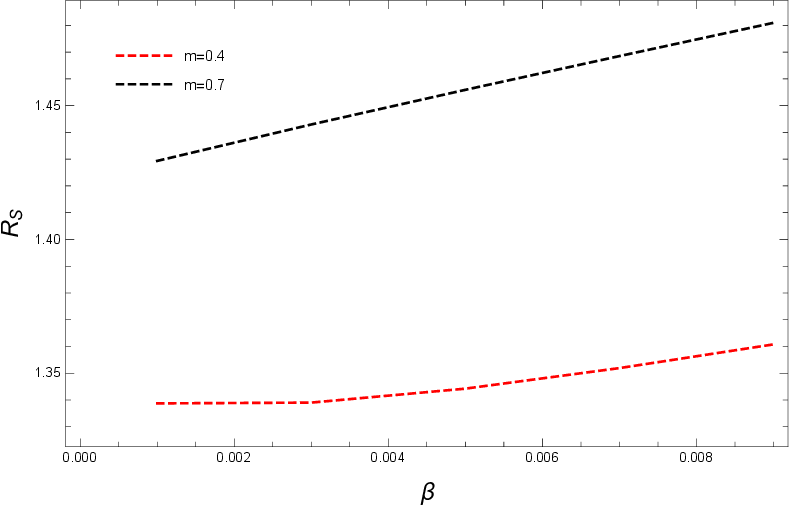}
         \subcaption{ $p=0.05$.}\label{fig:Rc}
   \end{minipage}
\caption{Correlation between the parameters and the shadow radius.}
\label{figrad}
\end{figure}

{Next}, in Fig. \ref{figshadw} we observe the impact of the $\beta$ parameter on the size of the black hole shadow. However, this {effect} seems to be relatively small compared with the other parameters' impact. { As the final visualization of this section,}  we present the close relationships between the shadow radius and the parameters in Fig. \ref{figrad}.

{Now, we explore how the Event Horizon Telescope (EHT)
measurements of M$87^{\ast }$ \cite{M871} and Sgr. $A^{\ast }$ \cite{SagA1} constrain the black hole parameters. To this end, we consider the necessary observational information tabulated in Table \ref{newtable1}.}
{\begin{table}[htb!]
\center%
\begin{tabular}{l|l|l|l}
\hline
Black hole & Mass ($M\odot )$ & Angular diameter $\Theta (\mu as)$ & 
Distance $\left( Mpc\right) $ \\ \hline\hline
Sgr. $A^{\ast }$ & $\left( 4.3\pm 0.013\right) 10^{6}$ & $\left( 48.7\pm
7\right) $ & $\left( 8.277\pm 0.033\right) 10^{-3}$ \\ \hline
M$87^{\ast }$ & $\left( 6.5\pm 0.90\right) 10^{9}$ & $\left( 42.7\pm
3\right) $ & $16.8$ \\ \hline\hline
\end{tabular}
\caption{Observational measurements of  Sgr. $A^{\ast }$ and M$87^{\ast }$    black holes. }\label{newtable1}
\end{table}}

\noindent{ Using this data, it is possible to determine the diameter of the shadow image of Sgr. $A^{\ast }$  with
\begin{equation}
d_{\text{Sgr.}A^{\ast }}=\frac{D\Theta }{M_{\text{Sgr.}A^{\ast }}}=\left(
9.5\pm 1.4\right) m,
\end{equation}%
where $d_{\text{Sgr.}A^{\ast }}=2R_{\text{Sgr.}A^{\ast }}.$ Thus, the radius of the observed shadow of the Sgr.$A^{\ast }$ in unit mass is
\begin{equation}
R_{\text{Sgr.}A^{\ast }}=\left( 4.75\pm 0.70\right) m.
\end{equation}
In the limit of  $p\rightarrow 0,$ the shadow radius reduces to 
\begin{equation}
R_{s}^{2}=\frac{6r_{p}^{2}}{3r_{p}-3m}+\frac{64\beta \left(
28080m^{4}-8505m^{3}r_{p}\right) }{15r^{7}\left( 3r_{p}-3m\right) ^{2}}.
\end{equation}%
So that we can estimate  the shadow radius of the Sgr.$A^{\ast }$ and compare it  Schwarzschild-like black hole in Einstein-Bel-Robinson gravity as tabulated in Table \ref{newtable2}:
\begin{table}[htb!]
\center%
\begin{tabular}{l|l}
\hline
$\beta $ & $R_{s}$ \\ \hline\hline
$10^{-2}$ & 2.53637 \\ \hline
$2\times 10^{-2}$ & 2.84698 \\ \hline
$5\times 10^{-2}$ & 3.31671 \\ \hline
$10^{-1}$ & 3.72288 \\ \hline
$2\times 10^{-1}$ & 4.17879 \\ \hline
$5\times 10^{-1}$ & 4.86826 \\ \hline\hline
\end{tabular}
\caption{Shadow radius comparison of the  Sgr.$A^{\ast }$ black hole with the Schwarzschild-like black hole in EBR gravity. }\label{newtable2}
\end{table}

\newpage
The results given in Table \ref{newtable2} show that the shadow radius of the black hole increases with the increasing value of the EBR gravity parameter.

As a final remark, we have to mention that other studies in the literature argue that these observations may be compatible with other alternative models to the black hole scenario as well \cite{Jao2, Jao3, Jao4}. We refrain from giving further details in this regard here.  }

\section{Energy emission rate} \label{sec3}

\begin{figure}[htb!]
\begin{minipage}[t]{0.5\textwidth}
        \centering
        \includegraphics[width=\textwidth]{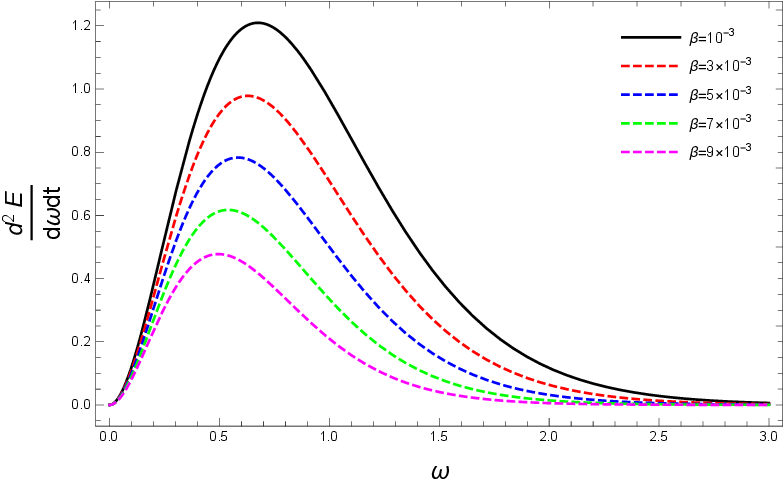}

       \subcaption{ $m=0.4$, and $p=0.05$.}\label{fig:ema}
   \end{minipage}%
\begin{minipage}[t]{0.5\textwidth}
        \centering
        \includegraphics[width=\textwidth]{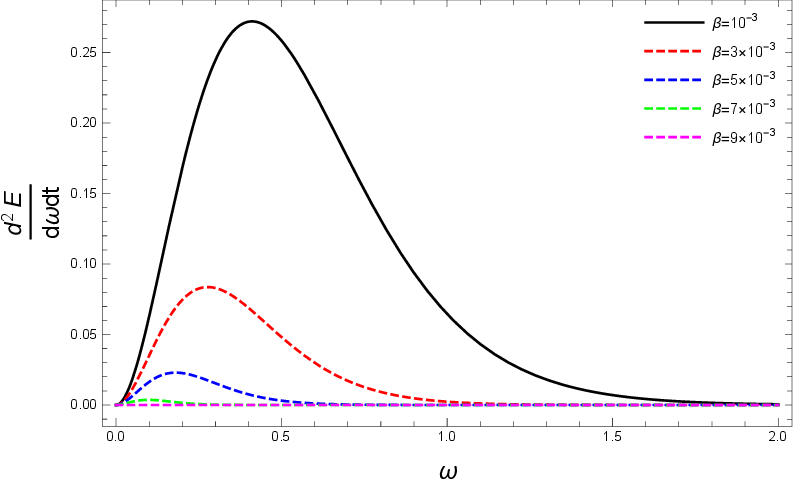}
         \subcaption{ $m=0.7$, and $p=0.05$.}\label{fig:emb}
   \end{minipage}
\caption{Energy emission rate behaviors of the ERB black hole for various values of the $\beta$ parameter.}
\label{figem}
\end{figure}

Black holes can emit radiation through the Hawking radiation process. At high-energy conditions, the absorption cross-section typically fluctuates around a constant value called $\sigma _{\lim }$. For a very long distant observer, the absorption cross-section gradually approaches the region where the black hole's shadow is formed \cite{Wei}. For a nearly spherically symmetric black hole $\sigma _{\lim }$ is approximately assumed to be equal to the photon sphere area
\begin{equation}
\sigma _{\lim }\sim \pi R_{s}^{2}.
\end{equation}
Thus, for the examined black hole, the complete form of the energy emission rate  reads:
\begin{equation}
\frac{d^{2}E\left( \omega \right) }{dt\, d\omega }=\frac{2\pi ^{2}\sigma _{\lim
}}{e^{\frac{\omega }{T_{H}}}-1},
\end{equation}%
where $\omega $ and $T_{H}$ denote the frequency of the photon and  Hawking temperature, respectively. In Fig. \ref{figem} we depict the black hole's energy emission rate as a function of photon frequency $\omega $ for various $\beta$ values. 
\newpage
Fig. \ref{figem} illustrates that as $\beta $ increases, the peak of the energy emission rate shifts to lower frequencies.

\section{Plasma effects to the Shadow in EBR gravity} \label{sec4}

In this section, we aim to present how the plasma background will affect the black hole shadow in the EBR gravity. The main motivation to focus on this question relies on the idea that in most cases black holes are surrounded by a medium that changes the geodesic of photons { \cite{Chowdhuri, Molla}}. We start with considering a plasma with a refractive index $n=n\left( x^{i},\omega\right) $. Here, $\omega $ denotes photon frequency. The existence of the background plasma modifies the Hamiltonian by the extra terms that appear in the geodesic equations. Consequently, these modifications affect the trajectories of particles and present an explicit frequency dependence.

Let us assume the effective energy of the particles inside the plasma medium as \cite{Synge}
\begin{equation}
\hbar \omega =p_{\alpha }u^{\alpha }, 
\end{equation}%
and then, we write the relation between plasma frequency and the 4-momentum of the photon with
\begin{equation}
n^{2}=1+\frac{p_{\alpha }p^{\alpha }}{\left( p_{\alpha }u^{\alpha }\right)
^{2}}.
\end{equation}
Next, by using the relationship between the refractive index and the plasma frequency, $\omega _{p}$, we express it as 
\begin{equation}
n^{2}=1-\left( \frac{\omega _{p}}{\omega }\right) ^{2}.
\end{equation}%
Then, we follow the authors of \cite{Rogers, Saha}, and use radial power law form to write the refractive index of plasma %
\begin{equation}
n\left( r\right) =\sqrt{1-\frac{\rho }{r}}.
\end{equation}%
Here, $\rho \geq 0$ is a constant. In the plasma background, the Hamilton-Jacobi equation becomes \cite{Synge}:
\begin{equation}
\frac{\partial }{\partial \tau }S=-\frac{1}{2}\left[ g^{\mu \nu }p_{\mu}p_{\nu }-\left( n^{2}-1\right) \left( p_{0}\sqrt{-g^{00}}\right) ^{2}\right],
\end{equation}%
and it leads to modifications in certain vacuum equations including the equations of motion of photons. In particular, the new set of null-geodesic equations read
\begin{eqnarray}
\dot{t}&=&\frac{n^{2}E}{f\left( r\right) }, \\
r^{2}\dot{r}&=&\sqrt{r^{4}n^{2}E^{2}-r^{2}f\left( r\right) \left( L^{2}+%
\mathcal{K}\right) } , \\
r^{2}\dot{\theta}&=&\sqrt{\mathcal{K}-L^{2}\cot ^{2}\theta } , \\
L&=&r^{2}\sin ^{2}\theta \dot{\phi} .
\end{eqnarray}%
In this case, the effective radial potential takes the form 
\begin{equation}
\Tilde{V}_{eff}\left( r\right) =\frac{f\left( r\right) }{r^{2}}\left( L^{2}+%
\mathcal{K}\right) -n^{2}E^{2},
\end{equation}%
and in the absence of a plasma background, it reduces to Eq. \eqref{Vef}.  After repeating the same method given above we obtain the unstable circular orbit conditions as 
\begin{equation}
\left. \left( n\left( r\right) rf^{\prime }\left( r\right) -2n\left(
r\right) f\left( r\right) -2rn^{\prime }\left( r\right) f\left( r\right)
\right) \right\vert _{r=r_{p}}=0.
\end{equation}%
Then, with the help of the impact parameters, we obtain
\begin{equation}
\eta +\zeta ^{2}=r_{p}^{2}\frac{n^{2}\left( r_{p}\right) }{f\left(
r_{p}\right) }.  \label{c3}
\end{equation}

With numerical methods, we compute the photon radius and $\eta+\zeta ^{2}$ for different values of $\left( m,p,\beta \right) $ in the presence of the plasma background, and We tabulate our results in Tables \ref{tabpla1}, \ref{tabpla2}, and \ref{tabpla3}.
\begin{table}[htb!]
\center%
\begin{tabular}{l|ll|ll}
\hline
\multirow{2}{*}{} & \multicolumn{2}{|c|}{$\rho =0.4$} &
\multicolumn{2}{|c}{$\rho =0.2$} \\ \hline\hline
$p$ & $r_{p}$ & $\eta +\zeta ^{2}$ & $r_{p}$ & $\eta +\zeta ^{2}$ \\ \hline
$10^{-3}$ & 2.9114 & 6.54089 & 2.95793 & 6.94591 \\
$3\times 10^{-3}$ & 2.88369 & 4.82238 & 2.94543 & 5.0969 \\
$5\times 10^{-3}$ & 2.85849 & 3.83137 & 2.93448 & 4.03125 \\
$7\times 10^{-3}$ & 2.83575 & 3.18559 & 2.92506 & 3.33746 \\
$9\times 10^{-3}$ & 2.81538 & 2.73075 & 2.91714 & 2.84935 \\ \hline\hline
\end{tabular}
\label{tabm1}
\caption{The computed values of photon radius and impact parameters with two different values of plasma background parameter with $\beta =10^{-2}$ and $m=1$.}
\label{tabpla1}
\end{table}
\begin{table}[htb!]
\center%
\begin{tabular}{l|ll|ll}
\hline
\multirow{2}{*}{} & \multicolumn{2}{|c|}{$\rho =0.4$}& \multicolumn{2}{|c}{$\rho =0.2$} \\ \hline\hline
$m$ & $r_{p}$ & $\eta +\zeta ^{2}$ & $r_{p}$ & $\eta +\zeta ^{2}$ \\ \hline
$0.1$ & 0.650461 & 0.450281 & 0.679367 & 0.789876 \\
$0.08$ & 0.599232 & 0.347518 & 0.629579 & 0.677837 \\
$0.06$ & 0.536295 & 0.234569 & 0.570532 & 0.559665 \\
$0.04$ & 0.446746 & 0.0873573 & 0.496155 & 0.428216 \\
$0.02$ & / & / & 0.389320 & 0.263556 \\ \hline\hline
\end{tabular}
\label{tabm2}
\caption{The computed values of photon radius and impact parameters with two different values of plasma background parameter with $\beta =10^{-2}$ and $p=0.005$.}
\label{tabpla2}
\end{table}
\begin{table}[htb!]
\center%
\begin{tabular}{l|ll|ll}
\hline
\multirow{2}{*}{} & \multicolumn{2}{|c|}{$\rho =0.4$}& \multicolumn{2}{|c}{$\rho =0.2$} \\ \hline\hline
$\beta $ & $r_{p}$ & $\eta +\zeta ^{2}$ & $r_{p}$ & $\eta +\zeta ^{2}$ \\
\hline
$10^{-3}$ & 1.13057 & 0.939047 & 1.22393 & 1.09423 \\
$3\times 10^{-3}$ & 1.27756 & 0.910117 & 1.36038 & 1.04595 \\
$5\times 10^{-3}$ & 1.3702 & 0.897703 & 1.44959 & 1.02233 \\
$7\times 10^{-3}$ & 1.44026 & 0.890983 & 1.51805 & 1.00805 \\
$9\times 10^{-3}$ & 1.49761 & 0.863585 & 1.5745 & 0.998692 \\ \hline\hline
\end{tabular}
\label{tabm3}
\caption{The computed values of photon radius and impact parameters with two different values of plasma background parameter with $m=0.4$ and $p=5\times 10^{-2}$.}
\label{tabpla3}
\end{table}

\begin{figure}[htb!]
\begin{minipage}[t]{0.5\textwidth}
        \centering
        \includegraphics[width=\textwidth]{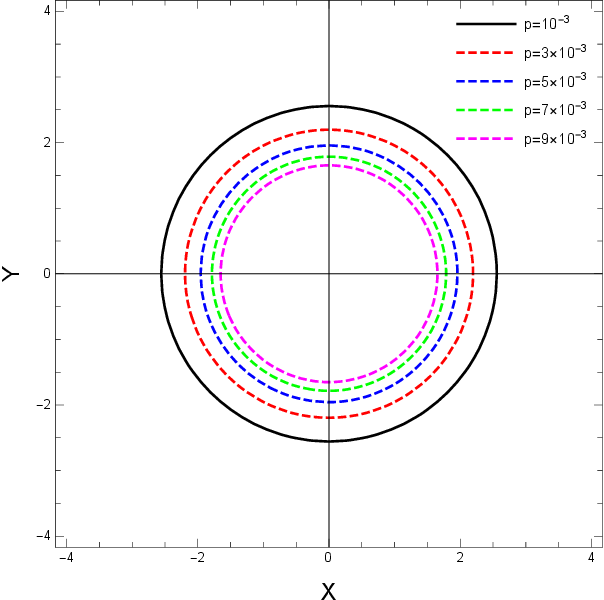}
       \subcaption{$\rho=0.4$.}\label{fig:psf}
   \end{minipage}%
\begin{minipage}[t]{0.5\textwidth}
        \centering
        \includegraphics[width=\textwidth]{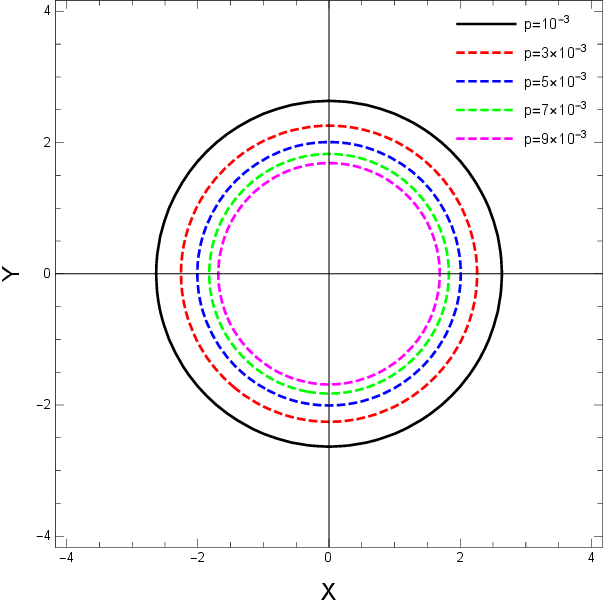}
         \subcaption{ $\rho=0.2$.}\label{fig:psg}
   \end{minipage}
\caption{Shadow shapes of the black holes in the equatorial plane by considering distinct values of $p$, for two different values of plasma medium and $m=1$, $\beta =10^{-2}.$}
\label{figppsh1}
\end{figure}

\begin{figure}[htb!]
\begin{minipage}[t]{0.5\textwidth}
        \centering
        \includegraphics[width=\textwidth]{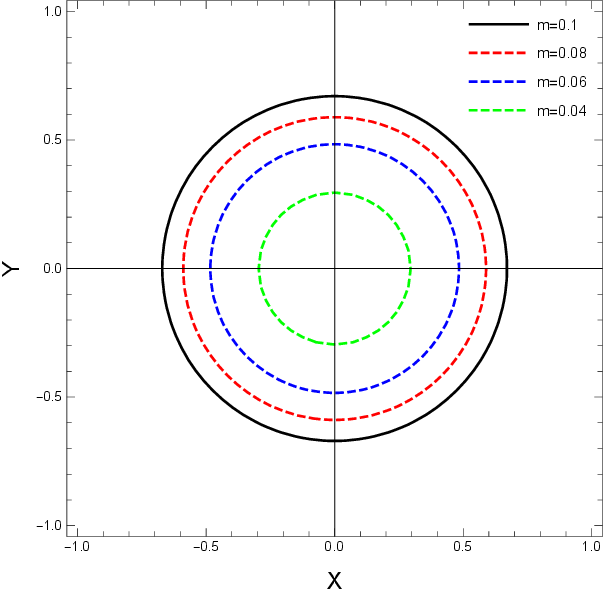}

       \subcaption{ $\rho=0.4$.}\label{fig:pph}
   \end{minipage}%
\begin{minipage}[t]{0.5\textwidth}
        \centering
        \includegraphics[width=\textwidth]{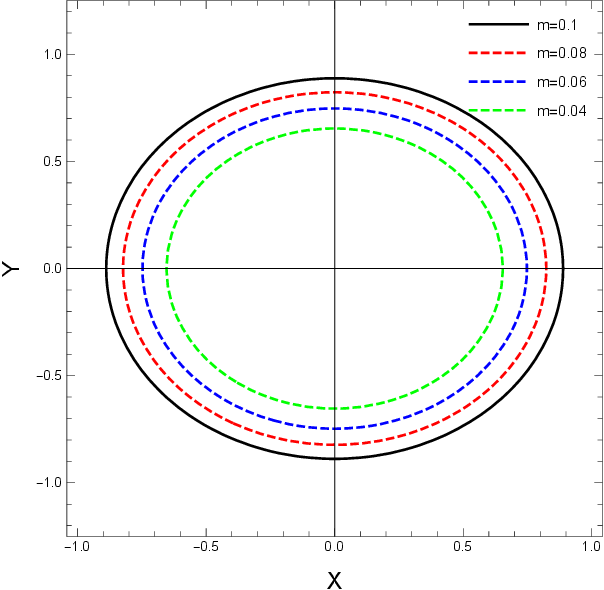}
         \subcaption{ $\rho=0.2$.}\label{fig:ppk}
   \end{minipage}
\caption{Shadow shapes of the black holes in the equatorial plane by considering distinct values of $m$, for two different values of plasma medium  and $%
p=5\times 10^{-3}$, $\beta =10^{-2}.$}
\label{figppsh2}
\end{figure}

\begin{figure}[htb!]
\begin{minipage}[t]{0.5\textwidth}
        \centering
        \includegraphics[width=\textwidth]{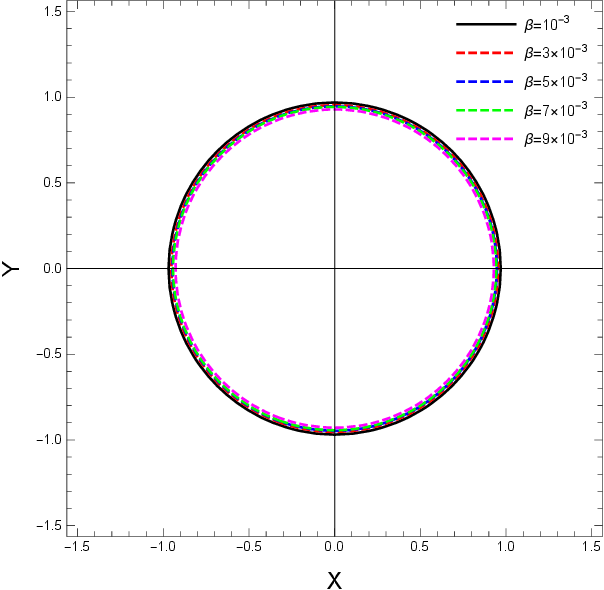}

       \subcaption{ $\rho=0.4$.}\label{fig:pplf}
   \end{minipage}%
\begin{minipage}[t]{0.5\textwidth}
        \centering
        \includegraphics[width=\textwidth]{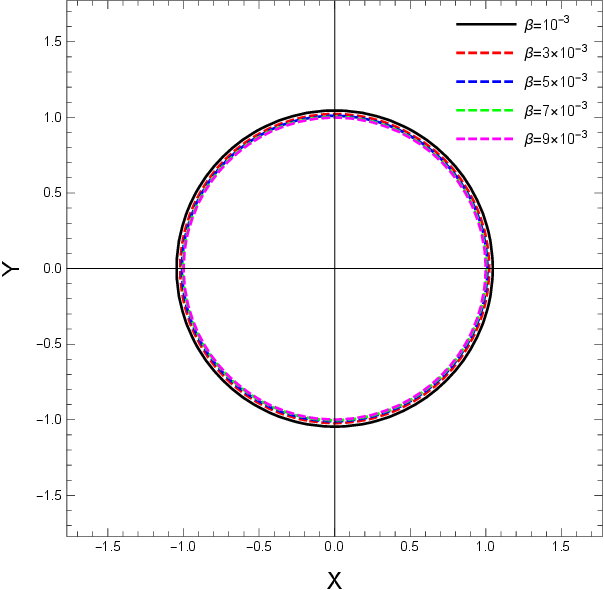}
         \subcaption{$\rho=0.2$.}\label{fig:pplg}
   \end{minipage}
\caption{Shadow shapes of the black holes in the equatorial plane by considering distinct values of $\beta $, for two different values of plasma medium and $p=5\times 10^{-2}$, $m=0.4$.}
\label{figppsh3}
\end{figure}

Next, we use the celestial coordinates 
\begin{equation}
Y=\pm \sqrt{\eta },  \label{c1}
\end{equation}%
\begin{equation}
X=-\zeta , \label{c2}
\end{equation}%
and we rewrite the shadow radius square in the presence of a plasma background as follows: 
\begin{equation}
X^{2}+Y^{2}=\zeta ^{2}+\eta =r_{p}^{2}\frac{n^{2}\left( r_{p}\right) }{%
f\left( r_{p}\right) }=R_{s}^{2}.
\end{equation}
Finally, we demonstrate the dependence of the shadow radius, $R_{s}$,  on the parameters $p$, $m$, $\beta$, and plasma parameters in Figs. \ref{figppsh1}-\ref{figppsh3}.

\newpage
\section{Conclusion}
In this manuscript, we investigated AdS black hole shadows in EBR gravity. To this end, we considered the AdS black hole solutions obtained by Sajadi et al very recently with perturbative analytic methods. After briefly showing how they found the line element in four-dimensional EBR gravity, we derived constants of motion from the Euler-Lagrange equation. Then, we used the Hamilton-Jacobi equation to find the null-geodesic equations of motion. With the separation method, we got the factorized relations and expressed the effective potential. Next, we recall the conditions for unstable null circular orbits and obtain the photon sphere radius. For arbitrary values of mass, pressure, and EBR gravity coupling constants, we calculated radii, and accordingly, we depicted the shadows. We found that an increase in pressure decreases the shadow radius. Unlike,  we observed that an increase in black hole mass increases the radius.  Moreover, we noted that for greater values of the gravity coupling constant, the shadow radius increases, however, we noticed that this impact is relatively smaller than the others. 
Next, we studied the energy emission rate. We found that the increase of the gravity coupling constant changes the peak value of the energy emission rate. Finally, we recalculated the geometry of AdS black hole shadows in the presence of a plasma background.

\section*{Acknowledgments}
{ The authors are thankful to the anonymous reviewer for the constructive comments.} This manuscript is supported by the Internal Project, [2023/2211], of Excellent Research of the Faculty of Science of Hradec Kr\'alov\'e University.

\section*{Data Availability Statements}

The authors declare that the data supporting the findings of this study are available within the article.

\end{document}